\documentclass[prd,preprintnumbers,twocolumn,amsmath,nofootinbib,amssymb]{revtex4}
\usepackage{graphicx,color,dcolumn,booktabs,bm}
\usepackage{longtable,lscape}
\usepackage{txfonts}
\usepackage{overpic}
\usepackage{amssymb}
\usepackage{epstopdf}
\usepackage{makecell}
\usepackage{indentfirst}
\usepackage{feynmf}   
\usepackage{slashed}  
\usepackage{cases}
\usepackage{color}
\usepackage{float}
\usepackage{multirow}
\usepackage{ulem}
\usepackage{epsfig,dsfont,amssymb,amsmath,amsfonts,amsbsy,mathrsfs}

\graphicspath{{Figures/}} %

\usepackage{hyperref}
\hypersetup{colorlinks,citecolor=blue,anchorcolor=red,menucolor=red, linkcolor=red,filecolor=red,runcolor=red,urlcolor=blue,frenchlinks=true}

\makeatletter
\@addtoreset{equation}{section}
\makeatother

\allowdisplaybreaks

\begin{document}

\title{Electromagnetic properties of possible triple-charm molecular hexaquarks}

\author{Chen-Ke Zhang$^{1,2,3,4}$}
\email{zhangchk2023@lzu.edu.cn}
\author{Fu-Lai Wang$^{1,2,3,4}$}
\email{wangfl2016@lzu.edu.cn}
\author{Si-Qiang Luo$^{1,2,3,4}$}
\email{luosq15@lzu.edu.cn}
\author{Xiang Liu$^{1,2,3,4}$\footnote{Corresponding author}}
\email{xiangliu@lzu.edu.cn}
\affiliation{$^1$School of Physical Science and Technology, Lanzhou University, Lanzhou 730000, China\\
$^2$Lanzhou Center for Theoretical Physics,
Key Laboratory of Theoretical Physics of Gansu Province,
Key Laboratory of Quantum Theory and Applications of MoE,
Gansu Provincial Research Center for Basic Disciplines of Quantum Physics, Lanzhou University, Lanzhou 730000, China\\
$^3$MoE Frontiers Science Center for Rare Isotopes, Lanzhou University, Lanzhou 730000, China\\
$^4$Research Center for Hadron and CSR Physics, Lanzhou University and Institute of Modern Physics of CAS, Lanzhou 730000, China}

\begin{abstract}
In this study, we investigate the radiative transitions of predicted triple-charm molecular hexaquarks, which play a significant role in understanding their overall spectroscopic properties. As experimentally measurable quantities, the radiative decay widths provide insights into the internal structures of these triple-charm molecular hexaquarks. Additionally, we calculate their corresponding magnetic moments, which, together with the radiative decay widths, offer a comprehensive picture of the electromagnetic properties of these exotic states. This information is valuable for guiding future experimental searches and advancing our understanding of these unique hadronic systems.
\end{abstract}
\maketitle

\section{Introduction}\label{sec1}

The search for exotic hadronic matter is a central focus in contemporary hadron spectroscopy, enriching our understanding of the hadron family and deepening insights into the nonperturbative dynamics of the strong interaction. Over the past two decades, significant progress has been made in the observation of exotic multiquark candidates, such as charmoniumlike $XYZ$ states, $P_c/P_{cs}$ pentaquark states, and the $T_{cc}(3875)^+$ state (see Refs.~\cite{Belle:2003nnu,LHCb:2015yax,LHCb:2019kea,LHCb:2021auc,LHCb:2021vvq,LHCb:2020jpq,LHCb:2022ogu,Liu:2013waa,Hosaka:2016pey,Chen:2016qju,Richard:2016eis,Olsen:2017bmm,Guo:2017jvc,Liu:2019zoy,Brambilla:2019esw,Meng:2022ozq,Chen:2022asf} for details).

\begin{figure}[htbp]
  \includegraphics[width=0.48\textwidth]{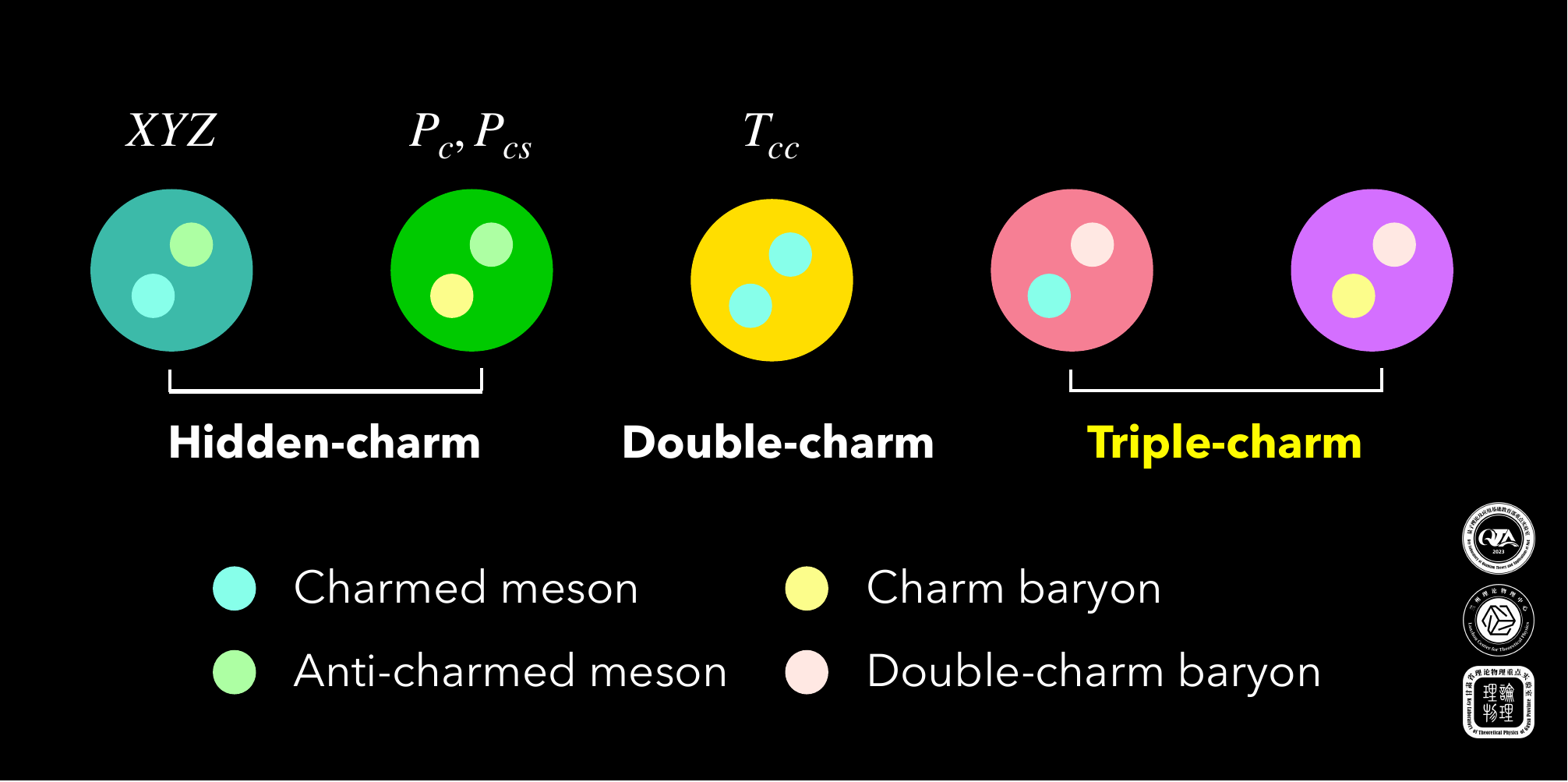}
  \caption{A revolution of hadronic molecular states with different charm numbers.}
  \label{revolution}
\end{figure}

As illustrated in Fig.~\ref{revolution}, various types of heavy-flavor hadronic molecular states have been identified. The observed charmoniumlike $XYZ$ states and $P_c/P_{cs}$ states are interpreted as hidden-charm molecular states, while the $T_{cc}(3875)^+$, discovered by the LHCb Collaboration, is classified as a double-charm tetraquark. Theoretical predictions by the Lanzhou group~\cite{Chen:2017jjn,Chen:2018pzd,Wang:2019aoc} suggest the existence of triple-charm hadronic molecular states, which have yet to be experimentally observed. The high-luminosity upgrade of the Large Hadron Collider (LHC)~\cite{LHCb:2018roe} offers a promising opportunity to search for these states.

In this work, we focus on triple-charm molecular hexaquarks, a novel class of exotic hadronic matter composed of a charm baryon and a double-charm baryon. The spectroscopic study of hadrons encompasses their mass spectrum, decay behavior, and production mechanisms. Among these, radiative transitions are a crucial aspect of decay behavior, as highlighted by the Particle Data Group (PDG)~\cite{ParticleDataGroup:2022pth}. Radiative transitions are not only experimentally measurable quantities but also provide valuable insights into the internal structures of both conventional and exotic hadrons. Therefore, the study of electromagnetic properties is essential for a comprehensive understanding of hadronic systems.

For the triple-charm molecular hexaquarks under discussion, we have previously calculated their mass spectrum in Ref.~\cite{Chen:2018pzd}. Building on this foundation, we now investigate their electromagnetic properties, which are critical for guiding experimental searches~\cite{Ozdem:2024yel,Mutuk:2024ltc,Guo:2023fih,Li:2024wxr,Li:2024jlq,Xu:2020flp,Meng:2021jnw}. Radiative transitions, as a key decay mode, are the primary focus of this study. Supported by the mass spectrum analysis, we calculate the radiative transitions for the predicted triple-charm molecular hexaquarks. Additionally, we determine their magnetic moments, which, together with the radiative decay widths, provide a complete picture of their electromagnetic properties.

This paper is organized as follows. After the introduction, Sec.~\ref{sec2} briefly reviews the predicted triple-charm molecular hexaquarks. In Sec.~\ref{sec3}, we present the radiative decays and magnetic moments of these states. Finally, Sec.~\ref{summary} provides a concise summary of our findings.

\section{A concise review for the predicted triple-charm molecular hexaquarks}\label{sec2}

With the observation of the double-charm baryon $\Xi_{cc}^{++}(3621)$ in the $\Lambda_c^+K^-\pi^+\pi^+$ invariant mass spectrum by the LHCb Collaboration~\cite{LHCb:2017iph}, the study of interactions between double-charm baryons and charmed hadrons (charmed mesons or charmed baryons) has garnered significant theoretical interest. A key objective in this field is to predict the existence of triple-charm molecular multiquark states, including triple-charm molecular pentaquarks~\cite{Wang:2019aoc,Chen:2017jjn} and hexaquarks~\cite{Chen:2018pzd}.

Focusing on the hexaquark system, the Lanzhou group systematically investigated the interactions between a double-charm baryon $\Xi_{cc}$ and $S$-wave charmed baryons ($\Lambda_c$, $\Sigma_c^{(*)}$, and $\Xi_{c}^{(\prime,*)}$) using the one-boson-exchange (OBE) model in 2018. Their results suggest the existence of several possible triple-charm molecular hexaquarks~\cite{Chen:2018pzd}. Following this, numerous theoretical studies have explored the properties of these states~\cite{Liu:2018zzu,Pan:2020xek,Pan:2019skd,Wang:2020jqu,Wang:2024riu,Chen:2024tuu,Junnarkar:2019equ}.

In Ref.~\cite{Liu:2018zzu}, the authors explored the connection between triple-charm molecular hexaquarks and hidden-charm molecular pentaquarks through heavy-antiquark-diquark symmetry (HADS). Using the OBE model, they studied the interactions between $\Xi_{cc}^{(*)}$ and $\Sigma_{c}^{(*)}$, constrained by the pentaquark system, and predicted several triple-charm molecular hexaquarks. Similarly, Ref.~\cite{Wang:2020jqu} employed QCD sum rules to investigate scalar and axial vector $\Xi_{cc}\Sigma_{c}$ states, considering both molecular and two-baryon scattering states. Their findings further support the existence of triple-charm molecular hexaquarks.

However, these studies primarily focus on the mass spectrum of these systems, which represents only one aspect of their spectroscopic behavior. To achieve a more comprehensive understanding, it is essential to investigate the electromagnetic properties of these triple-charm molecular hexaquarks, which will be the main focus of the present work. Here, we need to mention that the Lanzhou group conducted a systematic investigation of the electromagnetic properties including the radiative decays and the magnetic moments of the triple-charm molecular pentaquarks comprising the double-charm baryon and the charmed meson within the constituent quark model in Ref. \cite{Lai:2024jfe}.

\renewcommand\tabcolsep{0.19cm}
\renewcommand{\arraystretch}{1.50}
\begin{table}[htbp]
\caption{The predicted triple-charm molecular hexaquarks and their channels referred to be $|^{2S+1}L_J\rangle$ \cite{Chen:2018pzd}. Here, $I$ is the isospin of the triple-charm molecular hexaquarks, and $S$, $L$, and $J$ are the quantum numbers of spin, orbital, and total angular momentum, respectively. }\label{state}
\begin{tabular*}{86mm}{
@{\extracolsep{\fill}}m{16mm}<\centering
@{\vrule width 0.75pt}m{26mm}<\centering
@{\vrule width 0.75pt}m{16mm}<\centering
@{\vrule width 0.75pt}m{26mm}<\centering
}
    \toprule[1.0pt]\toprule[1.0pt]
    \multicolumn{4}{c}{$\Xi_{cc}\Sigma_c$ system}\\
    \midrule[0.75pt]
    $I(J^P)$ & Channel & $I(J^P)$ & Channels \\
    \midrule[0.75pt]
    $\frac{1}{2}(0^+),\frac{3}{2}(0^+)$ & $\left|\sideset{^1}{_0}{\mathop{S}}\right\rangle$ &
    $\frac{1}{2}(1^+)$ & $\left|\sideset{^3}{_1}{\mathop{S}}\right\rangle$, $\left|\sideset{^3}{_1}{\mathop{D}}\right\rangle$ \\
    \midrule[1.0pt]
    \multicolumn{4}{c}{$\Xi_{cc}\Sigma_c^*$ system}\\
    \midrule[0.75pt]
    $I(J^P)$ & Channels & $I(J^P)$ & Channels \\
    \midrule[0.75pt]
    $\frac{1}{2}(1^+)$ & $\left|\sideset{^3}{_1}{\mathop{S}}\right\rangle$, $\left|\sideset{^3}{_1}{\mathop{D}}\right\rangle$, $\left|\sideset{^5}{_1}{\mathop{D}}\right\rangle$ &
    $\frac{1}{2}(2^+),\frac{3}{2}(2^+)$ & $\left|\sideset{^5}{_2}{\mathop{S}}\right\rangle$, $\left|\sideset{^3}{_2}{\mathop{D}}\right\rangle$, $\left|\sideset{^5}{_2}{\mathop{D}}\right\rangle$ \\
    \midrule[1.0pt]
    \multicolumn{4}{c}{$\Xi_{cc}\Xi_c$ system}\\
    \midrule[0.75pt]
    $I(J^P)$ & Channel & $I(J^P)$ & Channels \\
    \midrule[0.75pt]
    $0(0^+)$ &
    $\left|\sideset{^1}{_0}{\mathop{S}}\right\rangle$ &
    $0(1^+)$ &
    $\left|\sideset{^3}{_1}{\mathop{S}}\right\rangle$, $\left|\sideset{^3}{_1}{\mathop{D}}\right\rangle$ \\
    \midrule[1.0pt]
    \multicolumn{4}{c}{$\Xi_{cc}\Xi_c^\prime$ system}\\
    \midrule[0.75pt]
    $I(J^P)$ & Channel & $I(J^P)$ & Channels \\
    \midrule[0.75pt]
    $0(0^+),1(0^+)$ &
    $\left|\sideset{^1}{_0}{\mathop{S}}\right\rangle$ &
    $0(1^+)$ &
    $\left|\sideset{^3}{_1}{\mathop{S}}\right\rangle$, $\left|\sideset{^3}{_1}{\mathop{D}}\right\rangle$ \\
    \midrule[1.0pt]
    \multicolumn{4}{c}{$\Xi_{cc}\Xi_c^*$ system}\\
    \midrule[0.75pt]
    $I(J^P)$ & Channels & $I(J^P)$ & Channels \\
    \midrule[0.75pt]
    $0(1^+)$ &
    $\left|\sideset{^3}{_1}{\mathop{S}}\right\rangle$, $\left|\sideset{^3}{_1}{\mathop{D}}\right\rangle$, $\left|\sideset{^5}{_1}{\mathop{D}}\right\rangle$ &
    $0(2^+),1(2^+)$ &
    $\left|\sideset{^5}{_2}{\mathop{S}}\right\rangle$, $\left|\sideset{^3}{_2}{\mathop{D}}\right\rangle$, $\left|\sideset{^5}{_2}{\mathop{D}}\right\rangle$ \\
    \bottomrule[1.0pt] \bottomrule[1.0pt]
\end{tabular*}
\end{table}

\begin{figure*}[htbp]
\centering
\includegraphics[width=0.85\linewidth]{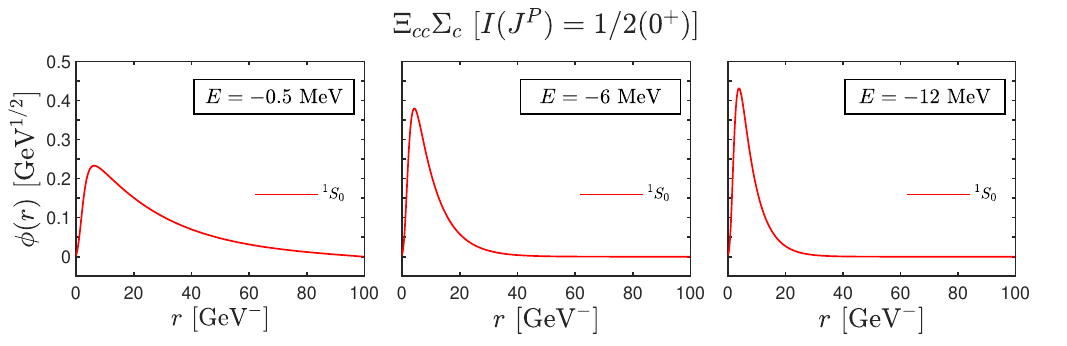}
\includegraphics[width=0.85\linewidth]{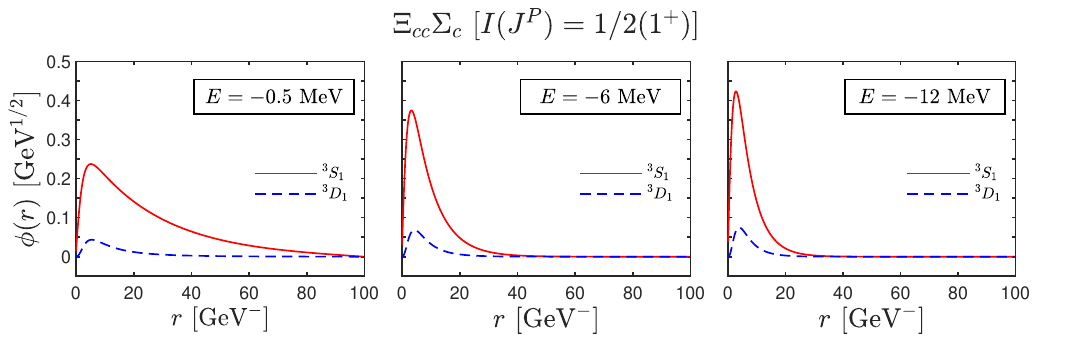}
\caption{The spatial wave functions of the $\Xi_{cc}\Sigma_c$ state with $I(J^P)=1/2(0^+)$ and the $\Xi_{cc}\Sigma_c$ state with $I(J^P)=1/2(1^+)$. Here, we also list the allowed $S$-wave and $D$-wave channels when considering the $S$-$D$ wave mixing effect. }
\label{spacewave}
\end{figure*}

Based on the Lanzhou group's study~\cite{Chen:2018pzd}, fourteen triple-charm molecular hexaquarks have been predicted. In this work, we concentrate on their electromagnetic properties. For convenience, Table~\ref{state} lists these predicted states. It should be emphasized that the spatial wave functions of the hadronic molecules constitute fundamental inputs for the calculation of their properties including the electromagnetic properties, which can be obtained from the quantitative study of their mass spectrum. Specifically, the spatial wave functions of the triple-charm molecular hexaquarks can be derived by solving the Schr\"odinger equation with the effective potentials obtained from the OBE model, as detailed in Ref.~\cite{Chen:2018pzd}. Notably, the binding properties of these molecular states play a significant role in shaping their spatial wave functions. Given the current absence of experimental data for these triple-charm hexaquarks \cite{ParticleDataGroup:2022pth}, we consider three typical binding energies of $-0.5$ MeV, $-6$ MeV, and $-12$ MeV to derive their spatial wave functions, which are then used to investigate their electromagnetic properties. For instance, the spatial wave functions of the $\Xi_{cc}\Sigma_c$ state with $I(J^P)=1/2(0^+)$ and the $\Xi_{cc}\Sigma_c$ state with $I(J^P)=1/2(1^+)$ can be obtained by solving the coupled-channel Schr\"odinger equation, as illustrated in Fig.~\ref{spacewave}. These numerically obtained spatial wave functions subsequently serve as critical inputs for the electromagnetic property calculations.

\section{Radiative decays and magnetic moments of the predicted triple-charm molecular hexaquarks}\label{sec3}

The electromagnetic properties of hadrons primarily include radiative decay widths and magnetic moments. In this section, we present the formalism for calculating these quantities within the constituent quark model, followed by a discussion of the numerical results for the radiative decays and the magnetic moments of the predicted triple-charm molecular hexaquarks.

For the radiative decay process $\mathcal{A} \rightarrow \mathcal{B}\gamma$, the general expression for the M1 radiative decay width is given by~\cite{Wang:2022nqs}
\begin{eqnarray}
    \Gamma_{\mathcal{A} \to \mathcal{B}\gamma}=
    \begin{cases}
        \frac{k^3}{3\pi}
        \frac{J_{\mathcal{A}}+1}{J_{\mathcal{A}}}
        \left|\mu_{\mathcal{A} \to \mathcal{B}}\right|^2    & \text{for}~~J_{\mathcal{A}}=J_{\mathcal{B}}, \\
        \frac{k^3}{3\pi}
        J_{\mathcal{A}}
        \left|\mu_{\mathcal{A} \to \mathcal{B}}\right|^2     & \text{for}~~J_{\mathcal{A}}=J_{\mathcal{B}}+1,\\
        \frac{k^3}{3\pi}
        \frac{J_{\mathcal{B}}(2J_{\mathcal{B}}+1)}{2J_{\mathcal{A}}+1}
        \left|\mu_{\mathcal{A} \to \mathcal{B}}\right|^2     & \text{for}~~J_{\mathcal{A}}=J_{\mathcal{B}}-1,
    \end{cases}
\end{eqnarray}
where $k={(m_{\mathcal{A}}^2-m_{\mathcal{B}}^2)}/{(2m_{\mathcal{A}})}$ is the momentum of the emitted photon, with $m_{\mathcal{A}}$ and $m_{\mathcal{B}}$ denoting the masses of the initial and final states, respectively. The symbols $J_{\mathcal{A}}$ and $J_{\mathcal{B}}$ represent the total angular momentum quantum numbers of the initial and final states, respectively. The transition magnetic moment, $\mu_{\mathcal{A} \to \mathcal{B}}$, is defined as \cite{Wang:2022nqs}
\begin{eqnarray}
    \mu_{\mathcal{A} \to \mathcal{B}}=
    \left\langle\psi_{J_{\mathcal{B}},J_{z}}\left
    |\sum_{j}\hat{\mu}_{zj}^{\rm spin}e^{-i {\bf k}\cdot{\bf r}_j}+\hat{\mu}_z^{\rm orbital}\right
    |\psi_{J_{\mathcal{A}},J_{z}}\right\rangle,\label{transitionmagnetic}
\end{eqnarray}
where $J_z={\rm min}\{J_{\mathcal{A}},J_{\mathcal{B}}\}$. The magnetic moment operators consist of two parts: the spin magnetic moment $\hat{\mu}_{zj}^{\rm spin}$ and the orbital magnetic moment $\hat{\mu}_z^{\rm orbital}$, expressed as
\begin{eqnarray}
    \hat{\mu}_{zj}^{\rm spin}&=&\frac{e_j}{2m_j}\hat{\sigma}_{zj},\label{mu_spin}\\
    \hat{\mu}_z^{\rm orbital}&=&\left(\frac{m_{h_1}}{m_{h_1}+m_{h_2}}\frac{e_{h_2}}{2m_{h_2}}+\frac{m_{h_{2}}}{m_{h_1}+m_{h_2}}\frac{e_{h_1}}{2m_{h_1}}\right)\hat{L}_z.\label{mu_orbital}
\end{eqnarray}
In Eq.~(\ref{mu_spin}), the spin magnetic moment operator acts on the wave functions of individual quarks, where $e_j$, $m_j$, and $\hat{\sigma}_{zj}$ are the charge, mass, and Pauli matrix of the $j$-th quark, respectively.
{In our analysis of the electromagnetic properties of the proposed hadronic molecular systems, we employ the constituent quark masses  $m_{u}=0.336\,\mathrm{GeV}$, $m_{d}=0.336\,\mathrm{GeV}$, $m_{s}=0.450\,\mathrm{GeV}$, and $m_{c}=1.680\,\mathrm{GeV}$, as established in Ref. \cite{Kumar:2005ei}. These parameter values have been extensively validated through previous studies on the magnetic moments of the hadronic molecular states \cite{Li:2021ryu,Zhou:2022gra,Wang:2022tib,Wang:2022nqs,Wang:2023aob,Wang:2023ael,Lai:2024jfe,Sheng:2024hkf,Wang:2024sbw,Wang:2024kke}.}
In Eq.~(\ref{mu_orbital}), the orbital magnetic moment operator acts on the wave functions of the hadrons $h_1$ and $h_2$ within the molecular state. Here, $h_1=\Xi_{cc}$ and $h_2=\Sigma_c^{(*)}/\Xi_c^{(\prime,*)}$ denote the double-charm and single-charm baryons, respectively. The quantities $e_{h_{1/2}}$ and $m_{h_{1/2}}$ represent the charge and mass of the hadrons, while $\hat{L}_z$ is the orbital angular momentum operator between hadrons $h_1$ and $h_2$.

In Eq.~(\ref{transitionmagnetic}), the term $e^{-i{\bf k} \cdot{\bf r}_j}$ represents the spatial wave function of the emitted photon, which can be expanded as
\begin{eqnarray}
    e^{-i{\bf k}\cdot{\bf r}_j}=\sum\limits_{l=0}^\infty\sum\limits_{m=-l}^l4\pi(-i)^lj_l(kr_j)Y_{lm}^*(\Omega_{\bf k})Y_{lm}(\Omega_{{\bf r}_j}),\label{photon0}
\end{eqnarray}
where $j_l(x)$ is the spherical Bessel function, and $Y_{l m}(\Omega_{\bf x})$ is the spherical harmonic function. For M1 radiative decays, the expansion is truncated at $l=0$ and $m=0$.

\renewcommand\tabcolsep{0.19cm}
\renewcommand{\arraystretch}{1.50}
\begin{table}[htbp]
\caption{The masses and the $\beta$ values of the single-charm baryons and the double-charm baryon. Here, the $\beta$ values are obtained from Refs.~\cite{Luo:2023sne,Lai:2024jfe,Yu:2022lel}, and the masses of baryons are taken as the average of its all isospin components sourced from the PDG~\cite{ParticleDataGroup:2022pth}.}\label{massbeta}
\begin{tabular*}{86mm}{
@{\extracolsep{\fill}}m{21mm}<\centering
@{\vrule width 0.75pt}m{21mm}<\centering
@{\vrule width 0.75pt}m{21mm}<\centering
@{\vrule width 0.75pt}m{21mm}<\centering
}
    \toprule[1.0pt]\toprule[1.0pt]
    Hadrons          & $m$ $({\rm GeV})$ & $\beta_\rho$ $({\rm GeV})$ & $\beta_\lambda$ $({\rm GeV})$ \\
    \midrule[0.75pt]
    $\Xi_c$          & $2.47$            & $0.301$                    & $0.383$                       \\
    $\Xi_c^{\prime}$ & $2.58$            & $0.252$                    & $0.383$                       \\
	$\Xi_c^{*}$      & $2.65$            & $0.243$                    & $0.358$                       \\
	$\Sigma_c$       & $2.45$            & $0.220$                    & $0.336$                       \\
    $\Sigma_c^{*}$   & $2.52$            & $0.212$                    & $0.315$                       \\
    $\Xi_{cc}$       & $3.62$            & $0.454$                    & $0.427$                       \\
    \bottomrule[1.0pt]\bottomrule[1.0pt]
\end{tabular*}
\end{table}

The total wave functions of the triple-charm molecular hexaquarks, $|\psi_{J_{\mathcal{A}},J_z}\rangle$ and $|\psi_{J_{\mathcal{B}},J_z}\rangle$, are composed of spatial, flavor, color, and spin components. While the flavor, color, and spin
wave functions are well-defined, the spatial wave functions
involve overlap integrals crucial for calculating radiative transitions. These spatial wave functions consist of two parts: one describing the internal structures of $h_1$ and $h_2$, and the other describing the interaction between $h_1$ and $h_2$. For baryons $h_1$ and $h_2$, the spatial wave functions can be decomposed into $\rho$-mode and $\lambda$-mode\footnote{In the double-charm baryon, the $\rho$-mode describes the relative position between the two charmed quarks, while the $\lambda$-mode describes the relative coordinate between the light quark and the center-of-mass of the two charmed quarks. Similarly, in the single-charm baryon, the $\rho$-mode describes the relative position between the two light quarks, and the $\lambda$-mode describes the relative coordinate between the charmed quark and the center-of-mass of the two light quarks.}. Each mode can be described by a simple harmonic oscillator (SHO) wave function:
\begin{eqnarray}
    \phi_{n,l,m}(\beta,{\mathbf r})
     &=&\sqrt{\frac{2n!}{\Gamma(n+l+\frac{3}{2})}}
    L_n^{l+\frac{1}{2}}(\beta^2 r^2)\beta^{l+\frac{3}{2}}\nonumber\\
    &&\times e^{-\frac{\beta^2r^2}{2}}r^lY_{lm}(\Omega_{\mathbf r}),
\end{eqnarray}
where $n$, $l$, and $m$ are the radial, orbital, and magnetic quantum numbers, respectively, and $\beta$ is a scaling parameter. The total spatial wave function of a baryon is then expressed as
\begin{equation}
\psi_{n_\rho,n_\lambda,l_\rho,l_\lambda,L,M}({\bm \rho},{\bm \lambda})=C_{l_\rho,m_\rho;l_\lambda,m_\lambda}^{LM}\phi_{n_\rho,l_\rho,m_\rho}(\beta_\rho,{\bm \rho})\phi_{n_\lambda,l_\lambda,m_\lambda}(\beta_\lambda,{\bm \lambda}),
\end{equation}
where $L$ and $M$ are the total orbital and magnetic quantum numbers of the baryon. Each baryon has two $\beta$ values, corresponding to the $\rho$-mode and $\lambda$-mode spatial wave functions. These values, along with the baryon masses, are provided in Table~\ref{massbeta}, taken from Refs.~\cite{Luo:2023sne,Lai:2024jfe,Yu:2022lel,ParticleDataGroup:2022pth}. {For the loosely hadronic molecular system, the spatial wave function deviates significantly from the SHO wave function. To account for this, we take the numerical spatial wave functions between two hadrons in the realistic calculations. Specifically, the numerical spatial wave functions of the triple-charm molecular hexaquarks can be obtained from the quantitative study of their mass spectrum by solving the coupled channel Sch\"{o}dinger equation based on the OBE effective potentials \cite{Chen:2018pzd}. Notably, the numerical spatial wave functions of the proposed triple-charm molecular hexaquarks depend on their binding energies. In the absence of experimental constraints for these exotic states, we adopt three representative binding energies: $-0.5$ MeV, $-6$ MeV, and $-12$ MeV, to systematically compute their numerical spatial wave functions. This approach allows us to evaluate the electromagnetic properties of these states under varying the binding energies.}

\renewcommand\tabcolsep{0.15cm}
\renewcommand{\arraystretch}{1.40}
\begin{table*}[htbp]
\centering
\caption{Transition magnetic moments and M1 radiative decay widths between the triple-charm molecular hexaquarks obtained from single-channel analysis and $S$-$D$ wave mixing analysis. Here, $\mu_{\cal A\to B}$ is given in units of $\mu_N$, where $\mu_N=\frac{e}{2m_p}$ is the nuclear magneton. For the radiative decays between the triple-charm molecular hexaquarks with different components, we assume identical binding energies for the initial and final states, using three typical values: $-0.5~{\rm MeV}$, $-6~{\rm MeV}$, and $-12~{\rm MeV}$. {For the radiative decays between the triple-charm molecular hexaquarks with the same component, we adopt three distinct combinations of the initial and final binding energies: ($-0.5~{\rm MeV}$, $-12~{\rm MeV}$), ($-0.5~{\rm MeV}$, $-6~{\rm MeV}$), and ($-6~{\rm MeV}$, $-12~{\rm MeV}$). In these combinations, the initial states are characterized by shallower binding energies, which correspond to higher system masses relative to the final states. Furthermore, due to the small radiative decay widths calculated for the transitions between the triple-charm molecular hexaquarks with the same component, only the combination yielding the largest decay width is explicitly tabulated in our results.}} \label{molecular_tmu_SD}
\begin{tabular}{
@{\extracolsep{\fill}}m{0.352\textwidth}<\centering
@{\vrule width 0.75pt}m{0.157\textwidth}<\centering
@{\vrule width 0.75pt}m{0.157\textwidth}<\centering
@{\vrule width 0.75pt}m{0.157\textwidth}<\centering
@{\vrule width 0.75pt}m{0.157\textwidth}<\centering
}
    \toprule[1pt]\toprule[1pt]
    \multirow{2}{*}{Radiative decays}
    & \multicolumn{2}{c@{\vrule width 0.75pt}}{Single channel analysis} & \multicolumn{2}{c}{$S$-$D$ wave mixing analysis} \\
    \Xcline{2-5}{0.75pt}
    & $\mu_{\mathcal{A} \rightarrow \mathcal{B}}$ & $\Gamma_{\mathcal{A} \rightarrow \mathcal{B} \gamma}$ $({\rm keV})$ & $\mu_{\mathcal{A} \rightarrow \mathcal{B}}$ & $\Gamma_{\mathcal{A} \rightarrow \mathcal{B} \gamma}$ $({\rm keV})$ \\
    \midrule[1.0pt]
    $\Xi_{cc}\Xi_c^\prime[0(1^+)]^{++}
    \rightarrow\Xi_{cc}\Xi_c[0(0^+)]^{++}\gamma$
	& $0.51,0.62,0.63$ & $0.93,1.34,1.38$ & $0.51,0.61,0.62$ & $0.92,1.32,1.36$ \\
	$\Xi_{cc}\Xi_c^\prime[0(1^+)]^{++
    }\rightarrow\Xi_{cc}\Xi_c[0(1^+)]^{++}\gamma$
    & $-0.51,-0.62,-0.63$ & $1.86,2.68,2.76$ & $-0.51,-0.61,-0.62$ & $1.84,2.64,2.72$ \\
	$\Xi_{cc}\Xi_c^\prime[1(0^+)]^{++}
    \rightarrow\Xi_{cc}\Xi_c[0(1^+)]^{++}\gamma$
	& $-0.62,-0.74,-0.74$ & $4.07,5.75,5.83$ & $-0.62,-0.73,-0.73$ & $4.05,5.61,5.64$ \\
	$\Xi_{cc}\Xi_c^\prime[0(0^+)]^{++}
    \rightarrow\Xi_{cc}\Xi_c[0(1^+)]^{++}\gamma$
	& $0.51,0.61,0.61$ & $2.70,3.87,3.89$ & $0.50,0.59,0.59$ & $2.67,3.70,3.64$ \\
	\midrule[0.75pt]
	$\Xi_{cc}\Xi_c^*[0(2^+)]^{++}
    \rightarrow\Xi_{cc}\Xi_c[0(1^+)]^{++}\gamma$
	& $0.48,0.68,0.71$ & $6.62,13.68,14.53$ & $0.47,0.68,0.70$ & $6.55,13.37,14.14$ \\
	$\Xi_{cc}\Xi_c^*[0(1^+)]^{++}
    \rightarrow\Xi_{cc}\Xi_c[0(0^+)]^{++}\gamma$
	& $0.54,0.58,0.80$ & $4.18,4.95,9.24$ & $0.53,0.46,0.77$ & $4.10,3.13,8.62$ \\
	$\Xi_{cc}\Xi_c^*[0(1^+)]^{++}
    \rightarrow\Xi_{cc}\Xi_c[0(1^+)]^{++}\gamma$
	& $0.27,0.29,0.40$ & $2.09,2.48,4.62$ & $0.27,0.23,0.38$ & $2.05,1.56,4.31$ \\
	$\Xi_{cc}\Xi_c^*[1(2^+)]^{++}
    \rightarrow\Xi_{cc}\Xi_c[0(1^+)]^{++}\gamma$
	& $-0.57,-0.81,-0.82$ & $9.45,18.97,19.84$ & $-0.57,-0.79,-0.80$ & $9.32,18.03,18.49$ \\
	\midrule[0.75pt]
	$\Xi_{cc}\Xi_c^*[0(2^+)]^{++}
    \rightarrow\Xi_{cc}\Xi_c^\prime[0(1^+)]^{++}\gamma$
	& $-0.36,-0.38,-0.38$ & $0.21,0.24,0.24$ & $-0.35,-0.38,-0.38$ & $0.21,0.23,0.23$ \\
	$\Xi_{cc}\Xi_c^*[0(1^+)]^{++}
    \rightarrow\Xi_{cc}\Xi_c^\prime[0(1^+)]^{++}\gamma$
	& $-0.20,-0.17,-0.21$ & $0.07,0.05,0.08$ & $-0.20,-0.14,-0.21$ & $0.07,0.03,0.07$ \\
	$\Xi_{cc}\Xi_c^*[0(1^+)]^{++}
    \rightarrow\Xi_{cc}\Xi_c^\prime[1(0^+)]^{++}\gamma$
	& $-0.58,-0.47,-0.60$ & $0.28,0.18,0.30$ & $-0.57,-0.35,-0.56$ & $0.27,0.10,0.26$ \\
	$\Xi_{cc}\Xi_c^*[0(1^+)]^{++}
    \rightarrow\Xi_{cc}\Xi_c^\prime[0(0^+)]^{++}\gamma$
	& $-0.41,-0.37,-0.43$ & $0.14,0.11,0.15$ & $-0.41,-0.31,-0.41$ & $0.14,0.08,0.14$ \\
	$\Xi_{cc}\Xi_c^*[1(2^+)]^{++}
    \rightarrow\Xi_{cc}\Xi_c^\prime[0(1^+)]^{++}\gamma$
	& $-0.51,-0.54,-0.54$ & $0.43,0.48,0.47$ & $-0.50,-0.52,-0.52$ & $0.42,0.45,0.44$ \\
	\midrule[0.75pt]
	$\Xi_{cc}\Xi_c[0(1^+)]^{++}
    \rightarrow\Xi_{cc}\Xi_c[0(0^+)]^{++}\gamma$
	& {$-0.02,-0.03,-0.03$} &
    {$<2.17\times10^{-6}$} &
    {$-0.02,-0.02,-0.03$} &
    {$<1.18\times10^{-6}$} \\
	\midrule[0.75pt]
	$\Xi_{cc}\Xi_c^\prime[1(0^+)]^{++}
    \rightarrow\Xi_{cc}\Xi_c^\prime[0(1^+)]^{++}\gamma$
	& {$0.36,0.39,0.46$} &
    {$<1.59\times10^{-3}$} &
    {$0.27,0.33,0.45$} &
    {$<9.36\times10^{-4}$} \\
	$\Xi_{cc}\Xi_c^\prime[0(0^+)]^{++}
    \rightarrow\Xi_{cc}\Xi_c^\prime[0(1^+)]^{++}\gamma$
	& {$0.41,0.47,0.56$} &
    {$<2.07\times10^{-3}$} &
    {$0.27,0.36,0.52$} &
    {$<8.94\times10^{-4}$} \\
	\midrule[0.75pt]
	$\Xi_{cc}\Xi_c^*[0(1^+)]^{++}
    \rightarrow\Xi_{cc}\Xi_c^*[0(2^+)]^{++}\gamma$
	& {$0.17,0.20,0.17$} &
    {$<4.27\times10^{-4}$} &
    {$0.12,0.16,0.12$} &
    {$<2.02\times10^{-4}$} \\
	$\Xi_{cc}\Xi_c^*[1(2^+)]^{++}
    \rightarrow\Xi_{cc}\Xi_c^*[0(2^+)]^{++}\gamma$
	& {$-1.41,-1.56,-1.79$} &
    {$<1.24\times10^{-2}$} &
    {$-1.06,1.30,-1.73$} &
    {$<7.09\times10^{-3}$} \\
    \midrule[1pt]
    $\Xi_{cc}\Sigma_c^*[1/2(2^+)]^{+++}
    \rightarrow\Xi_{cc}\Sigma_c[1/2(1^+)]^{+++}\gamma$
	& $0.74,0.79,0.79$ & $0.80,0.91,0.92$ & $0.73,0.77,0.77$ & $0.77,0.86,0.86$ \\
	$\Xi_{cc}\Sigma_c^*[1/2(2^+)]^{++}
    \rightarrow\Xi_{cc}\Sigma_c[1/2(1^+)]^{++}\gamma$
    & $-0.61,-0.65,-0.65$ & $0.54,0.62,0.62$ & $-0.60,-0.63,-0.63$ & $0.53,0.59,0.59$ \\
	$\Xi_{cc}\Sigma_c^*[1/2(1^+)]^{+++}
    \rightarrow\Xi_{cc}\Sigma_c[1/2(1^+)]^{+++}\gamma$
	& $0.42,0.44,0.44$ & $0.26,0.29,0.29$ & $0.41,0.42,0.41$ & $0.25,0.26,0.25$ \\
	$\Xi_{cc}\Sigma_c^*[1/2(1^+)]^{++}
    \rightarrow\Xi_{cc}\Sigma_c[1/2(1^+)]^{++}\gamma$
    & $-0.35,-0.37,-0.36$ & $0.18,0.20,0.19$ & $-0.34,-0.35,-0.34$ & $0.17,0.18,0.17$ \\
	$\Xi_{cc}\Sigma_c^*[1/2(1^+)]^{+++}
    \rightarrow\Xi_{cc}\Sigma_c[3/2(0^+)]^{+++}\gamma$
	& $-0.54,-0.56,-0.55$ & $0.22,0.23,0.23$ & $-0.53,-0.53,-0.51$ & $0.21,0.20,0.19$ \\
	$\Xi_{cc}\Sigma_c^*[1/2(1^+)]^{++}
    \rightarrow\Xi_{cc}\Sigma_c[3/2(0^+)]^{++}\gamma$
	& $-0.54,-0.56,-0.55$ & $0.22,0.23,0.23$ & $-0.53,-0.53,-0.51$ & $0.21,0.20,0.19$ \\
	$\Xi_{cc}\Sigma_c^*[1/2(1^+)]^{+++}
    \rightarrow\Xi_{cc}\Sigma_c[1/2(0^+)]^{+++}\gamma$
	& $0.84,0.90,0.88$ & $0.52,0.59,0.58$ & $0.83,0.86,0.83$ & $0.51,0.54,0.51$ \\
	$\Xi_{cc}\Sigma_c^*[1/2(1^+)]^{++}
    \rightarrow\Xi_{cc}\Sigma_c[1/2(0^+)]^{++}\gamma$
	& $-0.70,-0.74,-0.73$ & $0.36,0.40,0.39$ & $-0.69,-0.71,-0.69$ & $0.35,0.37,0.35$ \\
	$\Xi_{cc}\Sigma_c^*[3/2(2^+)]^{+++}
    \rightarrow\Xi_{cc}\Sigma_c[1/2(1^+)]^{+++}\gamma$
	& $-0.47,-0.49,-0.48$ & $0.32,0.35,0.34$ & $-0.45,-0.45,-0.44$ & $0.30,0.30,0.29$ \\
	$\Xi_{cc}\Sigma_c^*[3/2(2^+)]^{++}
    \rightarrow\Xi_{cc}\Sigma_c[1/2(1^+)]^{++}\gamma$
	& $-0.47,-0.49,-0.48$ & $0.32,0.35,0.34$ & $-0.45,-0.45,-0.44$ & $0.30,0.30,0.29$ \\
	\midrule[0.75pt]
	$\Xi_{cc}\Sigma_c[3/2(0^+)]^{+++}
    \rightarrow\Xi_{cc}\Sigma_c[1/2(1^+)]^{+++}\gamma$
	& {$0.34,0.37,0.43$} &
    {$<1.44\times10^{-3}$} &
    {$0.26,0.32,0.42$} &
    {$<8.60\times10^{-4}$} \\
	$\Xi_{cc}\Sigma_c[3/2(0^+)]^{++}
    \rightarrow\Xi_{cc}\Sigma_c[1/2(1^+)]^{++}\gamma$
	& {$0.34,0.37,0.43$} &
    {$<1.44\times10^{-3}$} &
    {$0.26,0.32,0.42$} &
    {$<8.60\times10^{-4}$} \\
	$\Xi_{cc}\Sigma_c[1/2(0^+)]^{+++}
    \rightarrow\Xi_{cc}\Sigma_c[1/2(1^+)]^{+++}\gamma$
	& {$-0.80,-0.94,-1.13$} &
    {$<8.07\times10^{-3}$} &
    {$-0.52,-0.71,-1.03$} &
    {$<3.37\times10^{-3}$} \\
	$\Xi_{cc}\Sigma_c[1/2(0^+)]^{++}
    \rightarrow\Xi_{cc}\Sigma_c[1/2(1^+)]^{++}\gamma$
	& {$0.60,0.70,0.85$} &
    {$<4.54\times10^{-3}$} &
    {$0.39,0.53,0.77$} &
    {$<1.90\times10^{-3}$} \\
	\midrule[0.75pt]
	$\Xi_{cc}\Sigma_c^*[1/2(1^+)]^{+++}
    \rightarrow\Xi_{cc}\Sigma_c^*[1/2(2^+)]^{+++}\gamma$
	& {$-0.33,-0.37,-0.45$} &
    {$<1.48\times10^{-3}$} &
    {$-0.22,-0.29,-0.41$} &
    {$<6.69\times10^{-4}$} \\
	$\Xi_{cc}\Sigma_c^*[1/2(1^+)]^{++}
    \rightarrow\Xi_{cc}\Sigma_c^*[1/2(2^+)]^{++}\gamma$
	& {$0.22,0.25,0.30$} &
    {$<6.59\times10^{-4}$} &
    {$0.15,0.19,0.27$} &
    {$<2.99\times10^{-4}$} \\
	$\Xi_{cc}\Sigma_c^*[3/2(2^+)]^{+++}
    \rightarrow\Xi_{cc}\Sigma_c^*[1/2(2^+)]^{+++}\gamma$
	& {$-1.32,-1.45,-1.63$} &
    {$<1.09\times10^{-2}$} &
    {$-0.99,-1.20,-1.52$} &
    {$<6.16\times10^{-3}$} \\
	$\Xi_{cc}\Sigma_c^*[3/2(2^+)]^{++}
    \rightarrow\Xi_{cc}\Sigma_c^*[1/2(2^+)]^{++}\gamma$
	& {$-1.32,-1.45,-1.63$} &
    {$<1.09\times10^{-2}$} &
    {$-0.99,-1.20,-1.52$} &
    {$<6.16\times10^{-3}$} \\
    \bottomrule[1pt]\bottomrule[1pt]
\end{tabular}
\end{table*}

In addition to radiative decay widths, the intrinsic magnetic moments are also crucial electromagnetic properties of hadrons. They are defined as
\begin{eqnarray}
    \mu_{\mathcal{A}}&=&\left\langle\psi_{J_{\mathcal{A}},J_{z}}\left|\sum_{j}\hat{\mu}_{zj}^{\rm spin}+\hat{\mu}_z^{\rm orbital}\right|\psi_{J_{\mathcal{A}},J_{z}}\right\rangle,\label{intrinsicmagneticmoment}
\end{eqnarray}
where $J_z$ is taken as the maximum $z$-component angular momentum of hadron, $J_z = J_{\mathcal{A}}$. The operators $\hat{\mu}_{zj}^{\rm spin}$ and $\hat{\mu}_z^{\rm orbital}$ are given in Eqs.~(\ref{mu_spin}) and (\ref{mu_orbital}), respectively. The intrinsic magnetic moment formula in Eq.~(\ref{intrinsicmagneticmoment}) is similar to the transition magnetic moment expression in Eq.~(\ref{transitionmagnetic}), but with two key differences: (i) the initial and final states in Eq.~(\ref{intrinsicmagneticmoment}) are identical, and (ii) there are no operators related to external photons, such as $e^{-i {\bf k}\cdot{\bf r}_j}$, in Eq.~(\ref{intrinsicmagneticmoment}).

Building on the established theoretical framework, we systematically calculate the transition magnetic moments and radiative decay widths for the triple-charm molecular hexaquark states. Table~\ref{molecular_tmu_SD} summarizes the numerical results for the transition magnetic moments and radiative decay widths for the triple-charm molecular hexaquark states, considering both single-channel and $S$-$D$ wave mixing schemes. It is noteworthy that the inclusion of the $S$-$D$ wave mixing effect results in only minor modifications to the transition magnetic moments and radiative decay widths for these discussed decay processes. This limited influence arises from the relatively small $D$-wave components introduced by the $S$-$D$ mixing effect \cite{Chen:2018pzd}, with the $S$-wave components remaining dominant in the formation of these loosely bound states.

From the numerical results presented in Table~\ref{molecular_tmu_SD}, three key observations can be drawn:
\begin{itemize}
  \item The radiative decay characteristics serve as valuable physical observables for probing the internal structures of the triple-charm molecular hexaquark states, highlighting their significance for future experimental research endeavors. A striking example is the substantial width difference between the $\Xi_{cc}\Xi_c^\prime[0(1^+)] \to \Xi_{cc}\Xi_c[0(0^+)]\gamma$ and $\Xi_{cc}\Xi_c^\prime[0(1^+)] \to \Xi_{cc}\Xi_c[0(1^+)]\gamma$ processes, which provides critical insights for determining the spin-parity quantum numbers of the $\Xi_{cc}\Xi_c$ system.
  \item The radiative decay widths for the triple-charm molecular hexaquark systems exhibit significant sensitivity to the third components of their isospin quantum numbers, reflecting the intrinsic relationship between the decay dynamics and the flavor wave function of both the initial and final states.
  \item {For the radiative decays between the triple-charm molecular hexaquarks with the same component, we take three distinct combinations of the initial and final binding energies: ($-0.5~{\rm MeV}$, $-12~{\rm MeV}$), ($-0.5~{\rm MeV}$, $-6~{\rm MeV}$), and ($-6~{\rm MeV}$, $-12~{\rm MeV}$). This can reflect the dependence of the transition magnetic moments and the M1 radiative decay widths between the proposed triple-charm molecular hexaquarks with the same component on the binding energies of the initial and final molecular states. While the calculated transition magnetic moments vary across these scenarios, our core conclusion remains robust, i.e., the radiative decays between the triple-charm molecular hexaquark systems with the same component exhibit remarkably smaller decay widths.} This feature can be attributed to the proximity of the masses of the initial and final triple-charm molecular hexaquark states, which significantly suppresses the kinetic phase space for these radiative transition processes.
\end{itemize}

{Here, we need to mention that we analyze the radiative decay behavior of the triple-charm molecular hexaquarks with different components under the assumption of identical binding energies for the initial and final molecular states in Table~\ref{molecular_tmu_SD}. In the following, we discuss the sensitivity of the radiative decay behavior of the proposed triple-charm molecular hexaquarks to different binding energies of the initial and final molecules. To quantify this dependence, we investigate the transition magnetic moment, the photon momentum, and the M1 radiative decay width of the $\Xi_{cc}\Xi_c^*[0(2^+)]^{++} \rightarrow\Xi_{cc}\Xi_c[0(1^+)]^{++}\gamma$ process when considering different binding energies of the initial and final molecules in the single-channel analysis, as detailed in Table~\ref{different_binding_energies}. This approach allows us to quantify how the different binding energies of the initial and final molecules influence the radiative decay behavior of the proposed triple-charm molecular hexaquarks. From Table~\ref{different_binding_energies}, our results demonstrate that the binding energies of the initial and final molecules can influence the M1 radiative decay width of the $\Xi_{cc}\Xi_c^*[0(2^+)]^{++} \rightarrow\Xi_{cc}\Xi_c[0(1^+)]^{++}\gamma$ process. Specifically, we see that the radiative decay width of the $\Xi_{cc}\Xi_c^*[0(2^+)]^{++} \rightarrow\Xi_{cc}\Xi_c[0(1^+)]^{++}\gamma$ process varies from several MeV to tens of MeV when the binding energies of the initial and final molecular states take different values between $-0.5\,{\rm MeV}$ and $-12\,{\rm MeV}$. Notably, this range aligns with predictions derived under the assumption of identical binding energies for the initial and final molecular states. Given the impact of the binding energies of the initial and final molecular states on the corresponding M1 radiative decay widths, we recommend the experimental determination of the binding energies of the proposed triple-charm molecular hexaquarks, which can advance our understanding of their inner structures and radiative decay behavior.}

\renewcommand\tabcolsep{0.001cm}
\renewcommand{\arraystretch}{1.50}
\begin{table}[htbp]
\caption{Transition magnetic moment, photon momentum, and M1 radiative decay width of the $\Xi_{cc}\Xi_c^*[0(2^+)]^{++} \rightarrow\Xi_{cc}\Xi_c[0(1^+)]^{++}\gamma$ process when considering different
binding energies of the initial and final molecules in the single-channel analysis.}
\label{different_binding_energies}
\begin{tabular*}{86mm}{
m{17mm}<\centering
@{\vrule width 0.75pt}m{17mm}<\centering
@{\vrule width 0.75pt}m{17mm}<\centering
@{\vrule width 0.75pt}m{17mm}<\centering
@{\vrule width 0.75pt}m{17mm}<\centering
}
    \toprule[1.0pt]\toprule[1.0pt]
    \multicolumn{2}{c@{\vrule width 0.75pt}}{Binding energy $({\rm MeV})$} &
    \multirow{2}{*}{$\mu_{\mathcal{A} \rightarrow \mathcal{B}}\,(\mu_N)$} &
    \multirow{2}{*}{$k$ $({\rm MeV})$} &
    \multirow{2}{*}{$\Gamma_{\mathcal{A} \rightarrow \mathcal{B}}\,({\rm keV})$} \\
    \Xcline{1-2}{0.75pt}
    Initial state& Final state&&\\
    \midrule[0.75pt]
    $-0.5$ & $-6$ & $0.53$ & $179.41$ & $8.87$ \\
    $-0.5$ & $-12$ & $0.49$ & $185.23$ & $8.48$ \\
    $-6$ & $-0.5$ & $0.53$ & $168.72$ & $7.51$ \\
    $-6$ & $-12$ & $0.68$ & $179.89$ & $14.95$ \\
    $-12$ & $-0.5$ & $0.50$ & $162.88$ & $6.03$ \\
    $-12$ & $-6$ & $0.69$ & $168.23$ & $12.56$\\
    \bottomrule[1.0pt] \bottomrule[1.0pt]
\end{tabular*}
\end{table}

\renewcommand\tabcolsep{0.06cm}
\renewcommand{\arraystretch}{1.50}
\begin{table}[!htbp]
\caption{The magnetic moment properties of the triple-charm molecular hexaquarks obtained through the single channel analysis and the $S$-$D$ wave mixing analysis. Here, we take three typical binding energies of $-0.5~{\rm MeV}$, $-6~{\rm MeV}$, and $-12~{\rm MeV}$ for the molecular hexaquarks $\Xi_{cc}\Xi_c^{(\prime,*)}$ and $\Xi_{cc}\Sigma_c^{(*)}$ to present their magnetic moment properties when considering the $S$-$D$ wave mixing effect.
}\label{molecular_mu_SD}
\begin{tabular}{
                      m{30mm}<\centering
@{\vrule width 0.75pt}m{27mm}<\centering
@{\vrule width 0.75pt}m{27mm}<\centering
}
    \toprule[1pt]\toprule[1pt]
    \multirow{2}{*}{Molecules} & \multicolumn{2}{c}{$\mu_H\, (\mu_N)$} \\
    \Xcline{2-3}{0.75pt}
    & Single channel & $S$-$D$ wave mixing \\
    \midrule[1.0pt]
	$\Xi_{cc}\Xi_c[0(1^+)]^{++}$ & $0.71$ & $0.71,0.71,0.71$ \\
	\midrule[0.75pt]
	$\Xi_{cc}\Xi_c^\prime[0(1^+)]^{++}$ & $0.06$ & $0.06,0.06,0.06$ \\
	\midrule[0.75pt]
	$\Xi_{cc}\Xi_c^*[0(1^+)]^{++}$ & $-0.05$ & $-0.05,-0.05,-0.05$ \\
	\midrule[0.75pt]
	$\Xi_{cc}\Xi_c^*[0(2^+)]^{++}$ & $0.48$ & $0.48,0.47,0.47$ \\
	\midrule[0.75pt]
	$\Xi_{cc}\Xi_c^*[1(2^+)]^{+++}$
	& $1.41$ & $1.41,1.41,1.40$ \\
	$\Xi_{cc}\Xi_c^*[1(2^+)]^{++}$
	& $0.48$ & $0.48,0.47,0.47$ \\
	$\Xi_{cc}\Xi_c^*[1(2^+)]^{+}$
	& $-0.45$ & $-0.45,-0.47,-0.47$ \\
	\midrule[0.75pt]
	$\Xi_{cc}\Sigma_c[1/2(1^+)]^{+++}$
	& $2.23$ & $2.18,2.14,2.14$ \\
	$\Xi_{cc}\Sigma_c[1/2(1^+)]^{++}$
	& $-0.56$ & $-0.54,-0.53,-0.53$ \\
	\midrule[0.75pt]
	$\Xi_{cc}\Sigma_c^*[1/2(1^+)]^{+++}$
	& $2.39$ & $2.36,2.30,2.27$ \\
	$\Xi_{cc}\Sigma_c^*[1/2(1^+)]^{++}$
	& $-0.56$ & $-0.55,-0.54,-0.53$ \\
	\midrule[0.75pt]
	$\Xi_{cc}\Sigma_c^*[1/2(2^+)]^{+++}$
	& $3.66$ & $3.63,3.61,3.60$ \\
	$\Xi_{cc}\Sigma_c^*[1/2(2^+)]^{++}$
	& $-0.37$ & $-0.37,-0.37,-0.37$ \\
	\midrule[0.75pt]
    $\Xi_{cc}\Sigma_c^*[3/2(2^+)]^{++++}$
	& $3.97$ & $3.93,3.87,3.86$ \\
	$\Xi_{cc}\Sigma_c^*[3/2(2^+)]^{+++}$
    & $2.42$ & $2.38,2.34,2.33$ \\
	$\Xi_{cc}\Sigma_c^*[3/2(2^+)]^{++}$
    & $0.87$ & $0.84,0.80,0.80$ \\
	$\Xi_{cc}\Sigma_c^*[3/2(2^+)]^{+}$
	& $-0.68$ & $-0.70,-0.73,-0.74$ \\
    \bottomrule[1pt]\bottomrule[1pt]
\end{tabular}
\end{table}

In addition to investigating the radiative decay widths for the triple-charm molecular hexaquark systems, we also discuss the intrinsic magnetic moments for the triple-charm molecular hexaquarks. In Table~\ref{molecular_mu_SD}, we provide the intrinsic magnetic moments for the triple-charm molecular hexaquarks. Similar to the investigation of the radiative decay widths for the triple-charm molecular hexaquark systems, we account for the influences of the $S$-$D$ wave mixing effect in the study of their magnetic moments. {In the single channel analysis, the magnetic moments of the proposed triple-charm molecular hexaquarks do not depend on the corresponding binding energies. This independence arises because the overlap of the spatial wave function of the $S$-wave component satisfies the normalization condition.}

As summarized in Table~\ref{molecular_mu_SD}, the magnetic moment serves as a crucial intrinsic property of the triple-charm molecular hexaquark systems, providing key insights into their internal structures. A notable illustration can be found in the $\Xi_{cc}\Xi_c^*$ system, where both the $\Xi_{cc}\Xi_c^*$ $I(J^P)=0(1^+)$ and $0(2^+)$ states share identical constituent particles (a double-charm baryon $\Xi_{cc}$ and a charm baryon $\Xi_c^*$), yet exhibit markedly different magnetic moment properties due to their distinct $J^P$ quantum numbers. This highlights how magnetic moments elucidate the coupling dynamics between a double-charm baryon $\Xi_{cc}$ and a charm baryon $\Xi_c^*$, serving as a valuable tool for distinguishing the spin-parity quantum numbers of the $\Xi_{cc}\Xi_c^*$ system in future experimental investigations. Furthermore, our analysis reveals that the triple-charm molecular hexaquarks with same $I(J^P)$ and different $I_3$ quantum numbers have different magnetic moments, since the triple-charm molecular hexaquarks with same $I(J^P)$ and different $I_3$ quantum numbers have different flavor wave functions. Similar to the case of the radiative decay widths for the triple-charm molecular hexaquark systems, the $D$-wave channels with the small contribution \cite{Chen:2018pzd} play a minor role to decorate the magnetic moment properties for the triple-charm molecular hexaquarks.

\section{Summary}\label{summary}

Since the discovery of the hidden-charm pentaquark states $P_c$ by the LHCb Collaboration~\cite{LHCb:2015yax,LHCb:2019kea}, molecular-type multiquark states have garnered significant attention within the hadron physics community. Identifying additional molecular-type multiquarks is a compelling research objective, as it not only contributes to the construction of ``Particle Zoo 2.0" but also provides critical insights into the nonperturbative dynamics of the strong interaction. The present work is part of this ongoing effort.

In this study, we focus on triple-charm molecular hexaquarks, a class of heavy-flavor molecular states. Previous theoretical predictions suggest that these states may exist in nature, with their mass spectrum already calculated~\cite{Chen:2018pzd}. While the mass spectrum is essential for their identification, it is not sufficient on its own. To provide a more comprehensive understanding, we investigate the electromagnetic properties of these predicted triple-charm molecular hexaquarks, including their transition magnetic moments, radiative decays, and magnetic moments. We hope that our results will inspire experimental efforts to search for these exotic states.

The exploration of triple-charm molecular hexaquarks represents a promising direction in the study of heavy-flavor molecular states. Our findings lay the groundwork for future experimental searches, potentially leading to the discovery of these exotic particles. By analyzing their electromagnetic properties in detail, we not only facilitate their identification but also gain deeper insights into the fundamental physics governing their interactions.

Looking ahead, future research should focus on refining theoretical models and conducting high-energy experiments to detect these hexaquarks. Close collaboration between theorists and experimentalists will be crucial in advancing our understanding of triple-charm molecular hexaquarks. Such synergy could significantly enhance our knowledge of the strong force and the behavior of multiquark systems. As experimental data continues to accumulate, we anticipate that the search for and study of triple-charm molecular hexaquarks will emerge as a vibrant and fruitful area of research in the field.

\begin{acknowledgments}

This work is supported by the National Natural Science Foundation of China under Grant Nos. 12335001, 12247155, 12247101, and 12405097, the China National Funds for Distinguished Young Scientists under Grant No. 11825503, National Key Research and Development Program of China under Contract No. 2020YFA0406400, the ‘111 Center’ under Grant No. B20063, the Natural Science Foundation of Gansu Province (No. 22JR5RA389, No. 25JRRA799), the fundamental Research Funds for the Central Universities, the project for top-notch innovative talents of Gansu province, and Talent Scientific Fund of Lanzhou University.

\end{acknowledgments}

\end{document}